\documentclass{article}
\usepackage{spconf,amsmath,amssymb,graphicx}
\usepackage{array,booktabs,etoolbox,textcomp}
\usepackage[table]{xcolor} 
\usepackage{url}
\usepackage{fontawesome5}
\usepackage[utf8]{inputenc}
\usepackage{hyperref} 

\definecolor{lightblue}{RGB}{224,247,250}

\title{Fusion Segment Transformer: Bi-directional attention guided fusion network for AI Generated Music Detection}

\name{Yumin Kim$^{*}$ and Seonghyeon Go$^{*}$\thanks{$^{*}$Equal contribution}}
\address{MIPPIA Inc. \\
\small \faGithub \hspace{0.5mm} \href{https://github.com/Mippia/ICASSP2026-FST?tab=readme-ov-file}{Code} \hspace{3mm} 
\faLink \hspace{0.5mm} \href{https://huggingface.co/spaces/mippia/AI-Music-Detection-FST}{Demo}}

\begin{document}
\maketitle

\begin{abstract}
With the rise of generative AI technology, anyone can now easily create and deploy AI-generated music, which has heightened the need for technical solutions to address copyright and ownership issues. While existing works mainly focused on short-audio, the challenge of full-audio detection, which requires modeling long-term structure and context, remains insufficiently explored. To address this, we propose an improved version of the Segment Transformer, termed the ~\textit{Fusion Segment Transformer}. As in our previous work, we extract content embeddings from short music segments using diverse feature extractors. Furthermore, we enhance the architecture for full-audio AI-generated music detection by introducing a Gated Fusion Layer that effectively integrates content and structural information, enabling the capture of long-term context. Experiments on the SONICS and AIME datasets show that our approach outperforms the previous model and recent baselines, achieving state-of-the-art results in AI-generated music detection. 
\end{abstract}


\section{Introduction}
\label{sec:intro}
In recent years, generative AI has shown significant progress in producing high-quality content. Music generation models are also developing in this way, which raises serious issues in the music industry~\cite{kaliakatsos2020artificial, chen2024applications}. Accessible AI music platforms now allow users to produce almost professional-level works across a range of musical aspects. End-to-end AI-generated music (AIGM) produced by these systems can synthesize almost every musical element, such as accompaniment, melody, lyrics, and production effects. Accordingly, the widespread availability of AI-generated music has posed a risk of intellectual property issues. In order to maintain the integrity of music copyrights, it has become important to develop a detection methodology that can distinguish between human-composed audio and AI creations.

Existing studies on AIGM detection have focused on extracting anomalies in various audio signals~\cite{li2024audio}. However, many existing methods are often limited to short audio sources like 5 or 30 seconds. Although some models have been applied to longer audio sources~\cite{rahman2024sonics}, they are designed for general audio signals and do not explicitly leverage the unique characteristics of music. Music is a long audio signal where the use of instruments and repetitive structures are prominent, depending on its tempo. For effective AIGM detection, it is important to consider these structural properties. In music theory, structural units such as bars and phrases are used to analyze musical composition. We refer to these musically meaningful units as ``Segments'', and suggest applying this concept to the AIGM detection task.

In our previous work, Segment Transformer~\cite{kim2025segment}, we employed a two-stage AIGM detection framework. Stage-1 extracted segment-level embeddings using various feature extractors. In Stage-2, we used the fact that the entire music is a sequence of segments to get the final classification result through a dual-pathway Transformer structure with a sequence of embeddings. However, we found that there is room to improve the Segment Transformer architecture. In particular, the Stage-2 fusion mechanism, simple concatenation, may fail to capture the relationships between content and structural information.

To this end, we propose the~\textit{Fusion Segment Transformer}, which follows a two-stage pipeline validated in previous work. We are additionally evaluating the Muffin Encoder~\cite{muffin} in Stage-1, which is specifically designed to address high-frequency artifacts in audio. In Stage-2, we incorporate bi-directional cross-attention and an adaptive gate fusion mechanism to capture the interaction between content and structural information better. Our model is explicitly designed to capture the long-term context within semantically connected audio segments, thereby enabling more reliable and fine-grained detection.
The main contributions of this work are:
(1) We introduce the Fusion Segment Transformer that integrates content and structural streams in music signals.  
(2) We flexibly integrate diverse feature extractors at the encoder structures, which enables comparative evaluation and complementary insights.
(3) We achieve state-of-the-art performance on the SONICS and AIME datasets, demonstrating the effectiveness of our fusion-based approach.  

\section{Preliminaries}
\label{sec:preliminaries}

\subsection{Segment Transformer}
We first provide a brief overview of the architecture of our previous work, the Segment Transformer~\cite{kim2025segment}. In Stage-1, we aim to extract meaningful embeddings from a given short audio segment. We use pre-trained self-supervised learning (SSL) models, including Wav2vec~\cite{baevski2020wav2vec}, Music2Vec~\cite{li2022map}, and MERT~\cite{li2023mert}, along with the pre-trained FXencoder~\cite{koo2023music}, to obtain embeddings that capture local musical characteristics. To better adapt these outputs to the AIGM detection task, we designed the AudioCAT framework, which flexibly employs feature extractor encoders and combines them with a fixed Cross-Attention–based Transformer decoder. This is followed by a classifier head for detection.
In Stage-2, we use a beat-tracking algorithm to split music tracks into four-bar segments and extract embeddings with the model in Stage-1, resulting in a sequence of segment embeddings $E = \{e_1, e_2, ..., e_N\}$ with each $e_i \in \mathbb{R}^d$, where $d$ is the embedding size of the extractor. We then compute a self-similarity matrix $M \in \mathbb{R}^{N \times N}$ to capture structural information~\cite{foote1999visualizing}, and we input both the embeddings and the similarity matrix into two parallel Transformer encoders. The final representation $h_{final} = h_{content} \oplus h_{similarity}$ integrates local segment features with global structural patterns by simple concatenation.

\section{Methods}
\label{sec:method}

This work follows a two-stage pipeline that extends the architecture of our previous work~\cite{kim2025segment}. Figure~\ref{fig:Stage1} shows the Stage-1 architecture for extracting embeddings from short audio segments. We employ feature extractors to obtain representations of short audio segments. As the setups for the other feature extractors in Figure~\ref{fig:Stage1}(a-d) are identical to those reported in our previous paper, we briefly describe the experimental methodology for the Muffin Encoder (e) in Section~\ref{sec:shortaudio}. We then present Stage-2 in Section~\ref{sec:fullaudio}, where we introduce our main contribution, an enhanced fusion-based architecture for full audio segment detection.

\subsection{Stage-1: Feature Embedding Extractor for Short Audio Segment Detection}
\label{sec:shortaudio}
In previous work, we employed various feature extractors with the AudioCAT to obtain segment-level embeddings, where models were trained using cross-entropy loss. But these models focus primarily on temporal patterns rather than detailed frequency-domain analysis. 

\begin{figure}[t]
\centering
\includegraphics[width=\columnwidth]{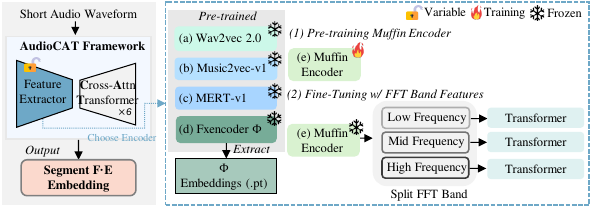}
\caption{The architecture of Stage-1 for short-audio segment detection. Various feature extractors (a–e) are flexibly selected on the Encoder of AudioCAT framework.}
\label{fig:Stage1}
\end{figure}

To explore whether frequency-domain features could complement AIGM detection~\cite{afchar2025fourier}, we experimented with integrating the Muffin Encoder~\cite{muffin} into the AudioCAT Encoder. It combines ResBlocks and a Multi-Receptive Field Fusion module to capture subtle tremors, irregular noise, and unnatural overtones in high-frequency ranges. We extract Mel-spectrograms from the 0–12 kHz band at 24 kHz and divide them into three multi-bands: low (0–2 kHz), mid (2–6 kHz), and high (6–12 kHz). We pre-train the Muffin Encoder with an MLP head and FFT-based band filtering, then freeze the Encoder and integrate it into the AudioCAT (Figure~\ref{fig:Stage1}(e)).

\subsection{Stage-2: Fusion Segment Transformer for Full Audio Segment Detection}
\label{sec:fullaudio}
Similar to our previous work~\cite{kim2025segment}, we prepare segment embeddings $E = \{e_1, e_2, ..., e_N\}$ from the whole music track using a beat-tracking algorithm and the segment feature extractor from Stage-1, where each $e_i \in \mathbb{R}^d$ and $d$ denotes the embedding dimension of the extractor. And we compute a self-similarity matrix, $SSM \in \mathbb{R}^{N \times N}$ as follows:
\begin{equation}
SSM_{ij} = \exp\left(-\frac{||e_i - e_j||^2}{d}\right)
\end{equation}

\begin{figure*}[t]
\centering
\includegraphics[width=\textwidth]{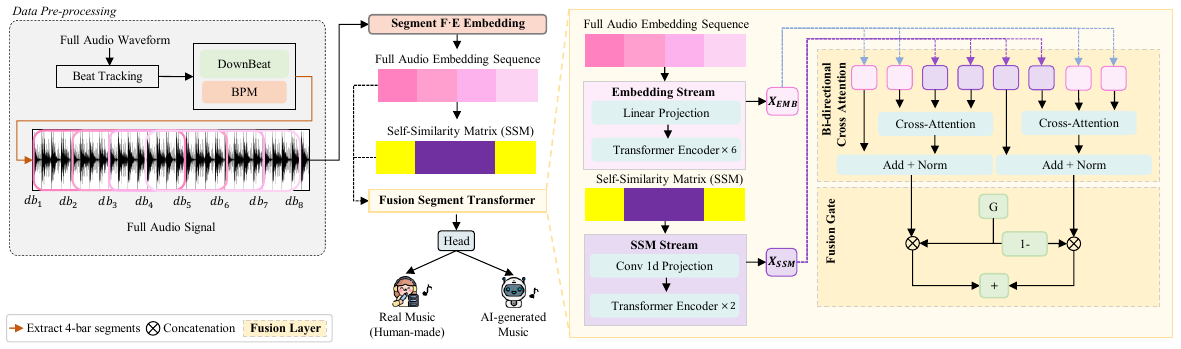}
\caption{Stage-2 fusion-enhanced Segment Transformer architecture with dual-stream processing, cross-modal fusion mechanism, and multi-scale adaptive pooling for full-audio AIGM detection.}
\label{fig:Stage2}
\end{figure*}

While our previous model, Segment Transformer, achieved competitive results through dual-pathway processing with simple concatenation, this approach treats content and structural information as separate features without considering how they interact. To address this limitation, we propose a fusion-enhanced architecture that enables better integration between \texttt{Embedding} and \texttt{SSM} streams through cross-stream attention mechanisms. Figure~\ref{fig:Stage2} shows our overall architecture, which includes dual-stream processing and a fusion mechanism for improved full-audio AIGM detection performance.

\subsubsection{Dual-Stream Processing}
We constructed an embedding stream to get features from the embedding itself obtained from Stage-1, and constructed an SSM stream to understand the music structural properties from the self-similarity of the embedding. The embedding stream is composed of a transformer encoder and is configured to focus on content features derived from Stage-1 embeddings that capture acoustic information. We used content features with positional encoding as an input, and feature $X_{EMB}$ can be obtained from $N$ segment embeddings. The SSM stream is configured to focus on information such as the repetitive structure that appears in music. The self-similarity matrix is processed as a 1D projection layer and then processed with the input of the transformer encoder layer together with positional encoding to obtain feature $X_{SSM}$.

\subsubsection{Cross-Modal Fusion Layer}
We used a cross-modal fusion layer structure to get comprehensive information from $X_{EMB}$ with contents-oriented embedding and $X_{SSM}$ with structure-oriented embedding. This layer combines two ideas: the use of bi-directional cross-attention to exchange information across modalities~\cite{jimale2025graph}, and the Gated Multimodal Unit~\cite{arevalo2017gated} to learn whether to focus on content or structure-based information. We obtained $X_{contents}$ by adding $X_{EMB}$ and the result of cross-attention, where $X_{EMB}$ serves as query and $X_{SSM}$ serves as key and value, followed by normalization. Similarly, we obtained $X_{structure}$ by adding $X_{SSM}$ and the result of cross-attention where $X_{SSM}$ serves as query and $X_{EMB}$ serves as key and value, followed by normalization.

The optimal integration of $X_{contents}$ and $X_{structure}$ is achieved through a learnable adaptive fusion gate:
\begin{equation}
X_{fused} = G \odot X_{contents} + (1 - G) \odot X_{structure}
\end{equation}
\begin{equation}
G = \sigma(W_g[X_{contents}; X_{structure}] + b_g)
\end{equation}

where the fusion gate $G$ is computed with learnable weight $W_g$ and bias $b_g$. The $X_{fused}$ embedding is fed into a classification head to determine whether the input is AIGM or human-composed.

\section{Experiments and Results}
\label{sec:expandresults}
\subsection{Datasets}
For training and evaluation of our proposed model, we used the SONICS~\cite{rahman2024sonics} dataset, which contains 48,090 real and 49,074 AI-generated tracks, each averaging about 176 seconds per track. Following the baseline setups, we resampled both real and fake music to 16kHz. SONICS provides a large amount of data, but requires unnecessary resampling for real tracks for fair condition (e.g., MERT, 44.1kHz → 16kHz → 24kHz). This may cause models to achieve misleadingly high performance by exploiting resampling artifacts rather than genuine musical content (See Section~\ref{results} for details). 

Furthermore, as noted in a recent studies~\cite{cros2025ai, sroka2025evaluating}, the high accuracy on SONICS is largely due to prominent generation artifacts from tools like Boomy\footnote{\url{https://boomy.com/}}, which simplify the classification task. To facilitate a more rigorous evaluation, we additionally evaluated our approach on the more challenging AIME dataset~\cite{grotschla2025benchmarking}, which contains more diverse generation sources with fewer obvious artifacts. The AIME dataset consists of 6,000 AI-generated tracks and 500 MTG-Jamendo real tracks, and we further collected additional real tracks from the MTG-Jamendo dataset, resulting in 6,000 real tracks in total. All datasets were split into training, validation, and test sets with a ratio of 8:1:1.

\subsection{Training Details and Evaluation Settings}
In Stage-1, we extracted 10-second segments from each track. Data augmentation included random pitch shifting (±2 semitones) and time stretching (0.8×–1.2×). SSL-based extractors and the FXencoder~\cite{koo2023music} followed prior configurations~\cite{kim2025segment}, while the Muffin Encoder~\cite{muffin} was trained in two stages: pre-training (lr=$1\times10^{-5}$, wd=$1\times10^{-6}$) and fine-tuning (lr=$2\times10^{-2}$, wd=$5\times10^{-2}$). All models were trained for 50 epochs with batch size 8 using BCE loss.

In Stage-2, tracks were segmented by length and downbeat, then padded or cropped to 48 segments before input to the fusion segment transformer. While 1--3 bar segmentation may benefit from data overlap, we found that 4-bar downbeat-based segmentation provides the most stable performance and allows consistent segment extraction even in free-tempo music. Downbeat information was extracted using Beat this!~\cite{foscarin2024beat} and quantized to a uniform temporal grid aligned with musical structure. This model was trained for 200 epochs with BCE loss using the Fused Adam optimizer ($lr=1\times10^{-4}$, $wd=1\times10^{-2}$) and early stopping.

Both stages were trained on a single NVIDIA RTX 5090 GPU. The model architecture is composed of 170.18M parameters in Stage-1 and 4.21M in Stage-2. Furthermore, our framework exhibits significant inference efficiency; it is capable of processing approximately 100,000 music tracks per day on a single RTX 5090 GPU, making it suitable for real-world, high-volume applications. 

We evaluated using Accuracy, Precision (Prec.), Recall (Sensitivity), F1-score, AUC (Area Under the Curve), and Specificity (Spec.).

\begin{table}[t]
  \centering
  \setlength{\tabcolsep}{4pt}
  \setlength{\arrayrulewidth}{0.6pt}
  \resizebox{\columnwidth}{!}{%
  \begin{tabular}{@{}lcccccc@{}}
    \toprule
    \multicolumn{7}{c}{\textbf{SONICS Dataset~\cite{rahman2024sonics} (Full Audio)}}\\
    \midrule
    \rowcolor{gray!10}
    \multicolumn{7}{c}{\textit{Baselines}~\cite{rahman2024sonics}}\\
    \textbf{Model} & ACC & Prec. & Recall & F1-Score & AUC & Spec.\\
    ConvNeXt~\cite{liu2022convnet}               & -- & -- & 0.95 & 0.96 & - & 0.98\\
    ViT~\cite{dosovitskiy2020image}                    & -- & -- & 0.82 & 0.89 & - & 0.98\\
    EfficientViT~\cite{cai2023efficientvit}           & -- & -- & 0.94 & 0.95 & - & 0.97\\
    SpecTTTra-$\gamma$~\cite{rahman2024sonics}    & -- & -- & 0.79 & 0.88 & - & \underline{0.99}\\
    SpecTTTra-$\beta$~\cite{rahman2024sonics}     & -- & -- & 0.86 & 0.92 & - & \underline{0.99}\\
    SpecTTTra-$\alpha$~\cite{rahman2024sonics}    & -- & -- & 0.96 & 0.97 & - & \underline{0.99}\\
    MERT (Standalone)   & \underline{0.9996} & \textbf{1.0000} & \underline{0.9996} & \underline{0.9996} & \textbf{1.0000} & \underline{0.9996} \\
    \rowcolor{lightblue}
    \multicolumn{7}{c}{\textit{Segment Transformer}~\cite{kim2025segment}}\\
    \textbf{Feature Extractor} \\
    Wav2vec 2.0  & 0.7473 & 0.7127 & 0.8726 & 0.7846 & 0.7412 & 0.6075\\
    Music2vec    & 0.9555 & 0.9609 & 0.9546 & 0.9577 & 0.9907 & 0.9566\\
    MERT         & 0.9992 & 0.9988 & 0.9996 & 0.9992 & \underline{0.9999} & 0.9987\\
    FXencoder    & 0.9955 & 0.9953 & 0.9961 & 0.9957 & 0.9995 & 0.9948\\
    \bottomrule
    \rowcolor{yellow!20}
\multicolumn{7}{c}{\textit{(Proposed) Fusion Segment Transformer}}\\
    Wav2vec 2.0   & 0.7758 & 0.7704 & 0.7568 & 0.7635 & 0.8563 & 0.7933 \\
    Music2vec     & 0.9994 & 0.9999 & 0.9988 & 0.9993 & \textbf{1.0000} & \underline{0.9999} \\
    MERT         & \textbf{0.9999} & \textbf{1.0000} & \textbf{0.9999} & \textbf{0.9999} &\underline{0.9999} & \textbf{1.0000} \\
    FXencoder     & 0.9974 & 0.9981 & 0.9964 & 0.9973 & 0.9998 & 0.9983 \\
    Muffin    & 0.9828 & 0.9863 & 0.9775 & 0.9819 & 0.9978 & 0.9876\\
    \bottomrule
  \end{tabular}}
  \caption{Comparison of various methods for full-audio detection (Stage-2) on SONICS dataset. Our proposed method is highlighted in yellow. Best: \textbf{Bold}; Second best: \underline{Underline}.}
  \label{tab:sonics_result}
\end{table}

\begin{table}[t]
  \centering
  \setlength{\tabcolsep}{4pt}
  \setlength{\arrayrulewidth}{0.6pt}
  \resizebox{\columnwidth}{!}{%
  \begin{tabular}{@{}lcccccc@{}}
    \toprule
    \multicolumn{7}{c}{\textbf{AIME Dataset~\cite{grotschla2025benchmarking} (Full Audio)}}\\
    \midrule
    \rowcolor{gray!10}
    \multicolumn{7}{c}{\textit{Baselines}}\\
    \textbf{Feature Extractor} & ACC & Prec. & Recall & F1-Score & AUC & Spec.\\
    MERT (Standalone)         & 0.9708 & 0.9750 & 0.9802 & \underline{0.9859} & 0.9939 & 0.9747\\
    \rowcolor{lightblue}
    \multicolumn{7}{c}{\textit{Segment Transformer}~\cite{kim2025segment}}\\
    Wav2vec 2.0  & 0.9082 & 0.8842 & 0.9375 & 0.9101 & 0.9642 & 0.8794\\
    Music2vec    & 0.9635 & 0.9560 & 0.9710 & 0.9635 & 0.9868 & 0.9561\\
    MERT         & \underline{0.9858} & \textbf{0.9917} & 0.9802 & \underline{0.9859} & \underline{0.9992} & \textbf{0.9916}\\
    FXencoder    & 0.9830 & 0.9843 & 0.9777 & 0.9810 & 0.9937 & {0.9873}\\
    Muffin       & 0.8739 & 0.8381 & 0.9241 & 0.8790 & 0.9213 & 0.8246\\
    \midrule
    \rowcolor{yellow!20}
    \multicolumn{7}{c}{\textit{(Proposed) Fusion Segment Transformer}}\\
    Wav2vec 2.0   & 0.9071 & 0.8760 & 0.9464 & 0.9099 & 0.9659 & 0.8684 \\
    Music2vec     & 0.9613 & 0.9579 & 0.9643 & 0.9611 & 0.9912 & 0.9583 \\
    MERT          & \textbf{0.9867} & \underline{0.9900} & \textbf{0.9835} & \textbf{0.9868} & \textbf{0.9995} & \underline{0.9899} \\
    FXencoder     & 0.9850 & 0.9843 & \underline{0.9821} & 0.9832 & {0.9971} & 0.9873 \\
    Muffin        & 0.8728 & 0.8377 & 0.9219 & 0.8778 & 0.9246 & 0.8246 \\
    \bottomrule
  \end{tabular}}%
  \caption{Comparison with the Segment Transformer for full-audio detection (Stage-2) on the AIME dataset. Our proposed method
is highlighted in yellow. Best: \textbf{Bold}; Second best: \underline{Underline}.}
  \label{tab:aime_result}
\end{table}

\subsection{Evaluation Results}
\label{results}
As shown in Tables~\ref{tab:sonics_result} and~\ref{tab:aime_result}, the proposed method consistently outperformed both the Segment Transformer~\cite{kim2025segment} and other state-of-the-art (SOTA) approaches on the SONICS~\cite{rahman2024sonics} dataset, and also surpassed the Segment Transformer on the AIME~\cite{grotschla2025benchmarking} dataset. Among feature extractors, the MERT-based model~\cite{li2023mert} achieved the best performance, suggesting that MERT's rich representations, pre-trained on a large-scale dataset, effectively capture key musical features. These results highlight the role of the cross-modal fusion layer, which is designed to focus less on the raw characteristics of the input data itself and more on the temporal relationships across musical segments. To further support the high performance of MERT on the SONICS, we also provide results from a standalone MERT model trained with a linear classifier head. Accordingly, when combined with extractors that already encode musically meaningful representations, such as FXencoder~\cite{koo2023music} or MERT, our model can effectively leverage inter-segment connections to achieve substantial accuracy. 

In contrast, Wav2vec 2.0~\cite{baevski2020wav2vec}, which is not specifically tailored for music-related features but for speeches, showed relatively lower performance than other extractors. Muffin Encoder~\cite{muffin}, despite its strength in frequency band-wise analysis, showed relatively lower performance compared to other extractors. This could be due to using only the encoder from an Muffin's adversarially trained architecture. However, frequency-domain analysis may still be beneficial for AIGM detection, requiring further research with refined architectures and training strategies. 

Therefore, the key contribution of this study lies in the robustness of the proposed fusion architecture itself, which consistently improved performance across nearly all evaluation metrics without relying on any specific feature extractor. This demonstrates that the proposed framework is extractor-agnostic and can be effectively applied to diverse input features.

\noindent \textbf{Statistical and Qualitative Analysis.} 
While the calculated $p$-value comparing the Fusion Segment Transformer (FST) and the Segment Transformer~\cite{kim2025segment} is approximately 0.09 ($> 0.05$), a closer inspection of the results reveals that our method primarily corrects predictions in the most challenging borderline cases. Specifically, by effectively leveraging long-term structural information through the gated fusion mechanism, FST reduces classification errors in scenarios involving high-quality AI-generated musics and real recordings with significant acoustic artifacts. This suggests that our model provides more robust decision boundaries in cases where content-only features may be ambiguous.

\subsection{Ablation Study}
\subsubsection{Impact of Segmentation Strategy}
We further investigated the necessity of musical structure-aware segmentation. When we replaced our 4-bar downbeat-based segmentation with a fixed-length approach (e.g., windows of 0--10s, 2.5--12.5s, and 5--15s), the validation accuracy of the Fusion Segment Transformer notably decreased from 0.9867 to 0.966. This performance drop suggests that fixed-length windowing fails to consistently capture the inherent structural patterns and rhythmic characteristics that are crucial for distinguishing between human-made and AI-generated music, especially in tracks with free-tempo elements.

\subsubsection{Effect of the Gated Fusion Layer}
\begin{figure}[t]
\centering
\includegraphics[width=\columnwidth]{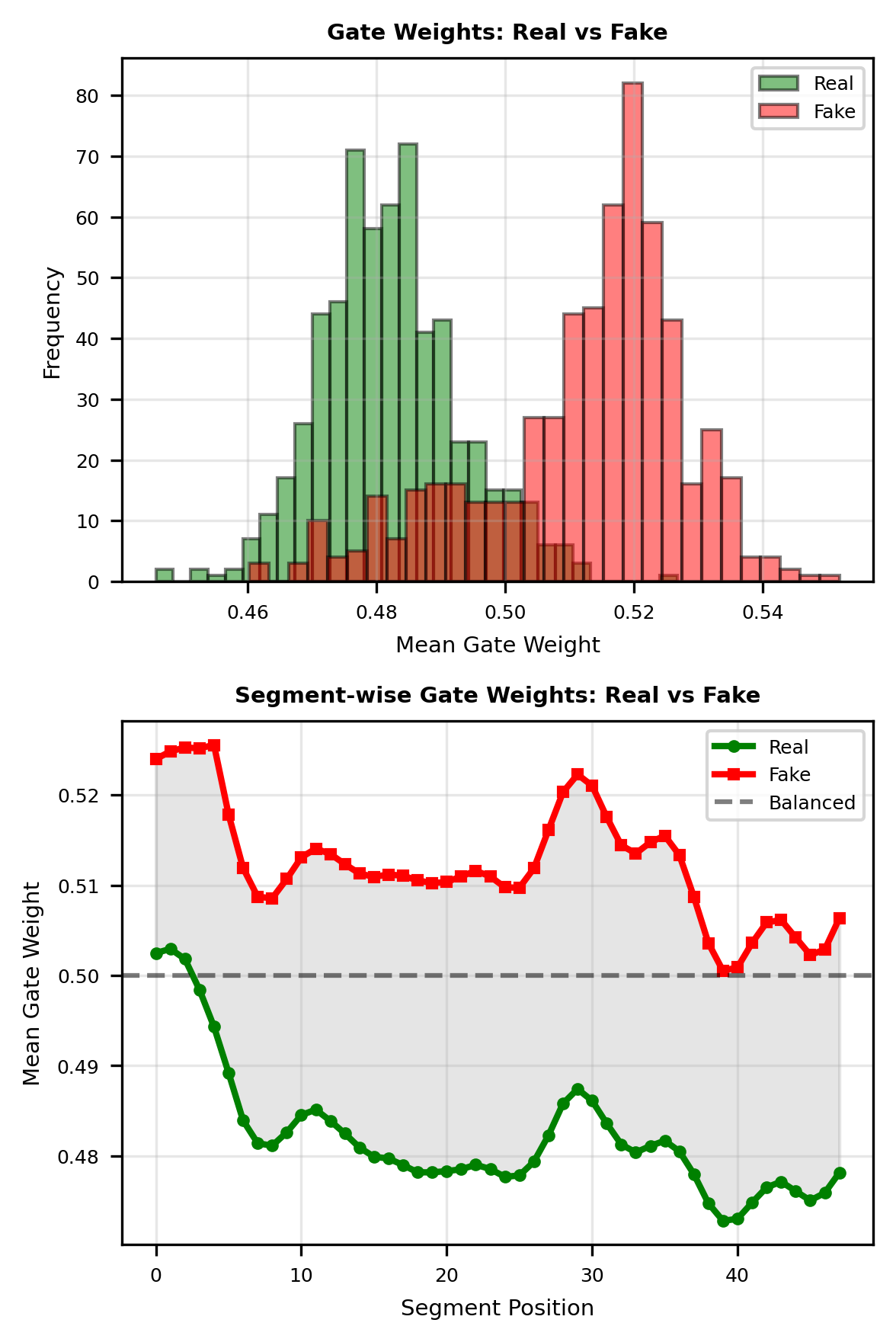}
\caption{Visualization of fusion gate dynamics for Real and Fake tracks on the AIME dataset~\cite{grotschla2025benchmarking}.}
\label{fig:fusion_gate_analysis}
\end{figure}

The primary advantage of the gated fusion layer lies in its ability to dynamically regulate the relative importance of content and structural representations across both temporal regions and target classes, rather than treating all interactions uniformly. As shown in Fig.~\ref{fig:fusion_gate_analysis}, the visualization of the fusion gate weights during inference demonstrates that the model assigns higher importance to content embeddings in the intro sections, where establishing the initial acoustic identity is critical. This adaptive weighting strategy enables the model to anchor long-term context formation on reliable local cues before progressively incorporating higher-level structural information. 

In addition, a clear class-dependent behavior is observed. Tracks classified as Real consistently assign higher weights to structural information, reflecting coherent global musical organization. In contrast, Fake tracks tend to rely more heavily on local content features, indicating weaker long-range structural consistency.

Such behavior cannot be replicated by uniform interaction mechanisms. When the gated fusion layer is replaced with a standard multi-head cross-attention module, a noticeable performance degradation is observed, with the MERT-based~\cite{li2023mert} accuracy dropping to 0.9850 on the AIME dataset~\cite{grotschla2025benchmarking}. These results demonstrate that the gated fusion layer plays a crucial role in selectively and stably integrating structural cues, which is essential for detecting subtle long-form irregularities in AI-generated music.

\section{Conclusions}
\label{sec:conclusions}
In this work, we present the Fusion Segment Transformer. Existing research has not fully exploited the structural characteristics of music. Even our previous work, the Segment Transformer, relied on simple concatenation in Stage-2, which was insufficient to capture the complex interactions between content and structural information. 
To address this limitation, the proposed model introduces an embedding stream and an SSM stream, and integrates them through bi-directional cross-attention and an adaptive gate fusion layer to achieve more effective fusion. As a result, experiments with various feature extractors consistently demonstrated performance improvements over existing methods. These findings indicate that a strategy that balances structural and content information is effective for AI-generated music detection. 

Future work could explore end-to-end model architectures and diverse encoder configurations to further enhance detection capabilities. Additionally, our results suggest the potential of a segment-based approach and music structure-aware fusion strategies in broader music information retrieval tasks.

\bibliographystyle{IEEEbib}
\bibliography{strings,refs}

@inproceedings{foote1999visualizing,
  title={Visualizing music and audio using self-similarity},
  author={Foote, Jonathan},
  booktitle={Proceedings of the seventh ACM international conference on Multimedia (Part 1)},
  pages={77--80},
  year={1999}
}

@article{cros2025ai,
  title={The AI Music Arms Race: On the Detection of AI-Generated Music},
  author={Cros Vila, Laura and Sturm, Bob and Casini, Luca and Dalmazzo, David},
  journal={Transactions of the International Society for Music Information Retrieval},
  volume={8},
  number={1},
  pages={179--194},
  year={2025}
}

@article{sroka2025evaluating,
  title={Evaluating fake music detection performance under audio augmentations},
  author={Sroka, Tomasz and Sidorczuk, Dominik and Modrzejewski, Mateusz and others},
  journal={arXiv preprint arXiv:2507.10447},
  year={2025}
}

@article{jimale2025graph,
  title={Graph-to-Text Generation with Bidirectional Dual Cross-Attention and Concatenation.},
  author={Jimale, Elias Lemuye and Chen, Wenyu and Al-antari, Mugahed A and Gu, Yeong Hyeon and Agbesi, Victor Kwaku and Feroze, Wasif and Akmel, Feidu and Assefa, Juhar Mohammed and Shahzad, Ali},
  journal={Mathematics (2227-7390)},
  volume={13},
  number={6},
  year={2025}
}

@article{arevalo2017gated,
  title={Gated multimodal units for information fusion},
  author={Arevalo, John and Solorio, Thamar and Montes-y-G{\'o}mez, Manuel and Gonz{\'a}lez, Fabio A},
  journal={arXiv preprint arXiv:1702.01992},
  year={2017}
}

@inproceedings{grotschla2025benchmarking,
  title={Benchmarking Music Generation Models and Metrics via Human Preference Studies},
  author={Gr{\"o}tschla, Florian and Solak, Ahmet and Lanzend{\"o}rfer, Luca A and Wattenhofer, Roger},
  booktitle={ICASSP 2025-2025 IEEE International Conference on Acoustics, Speech and Signal Processing (ICASSP)},
  pages={1--5},
  year={2025},
  organization={IEEE}
}

@article{rahman2024sonics,
  title={SONICS: Synthetic Or Not--Identifying Counterfeit Songs},
  author={Rahman, Md Awsafur and Hakim, Zaber Ibn Abdul and Sarker, Najibul Haque and Paul, Bishmoy and Fattah, Shaikh Anowarul},
  journal={arXiv preprint arXiv:2408.14080},
  year={2024}
}

@article{foscarin2024beat,
  title={Beat this! Accurate beat tracking without DBN postprocessing},
  author={Foscarin, Francesco and Schl{\"u}ter, Jan and Widmer, Gerhard},
  journal={arXiv preprint arXiv:2407.21658},
  year={2024}
}

@article{kaliakatsos2020artificial,
  title={Artificial intelligence methods for music generation: a review and future perspectives},
  author={Kaliakatsos-Papakostas, Maximos and Floros, Andreas and Vrahatis, Michael N},
  journal={Nature-inspired computation and swarm intelligence},
  pages={217--245},
  year={2020},
  publisher={Elsevier}
}

@article{chen2024applications,
  title={Applications and advances of artificial intelligence in music generation: A review},
  author={Chen, Yanxu and Huang, Linshu and Gou, Tian},
  journal={arXiv preprint arXiv:2409.03715},
  year={2024}
}

@article{li2024audio,
  title={From Audio Deepfake Detection to AI-Generated Music Detection--A Pathway and Overview},
  author={Li, Yupei and Milling, Manuel and Specia, Lucia and Schuller, Bj{\"o}rn W},
  journal={arXiv preprint arXiv:2412.00571},
  year={2024}
}

@inproceedings{koo2023music,
  title={Music mixing style transfer: A contrastive learning approach to disentangle audio effects},
  author={Koo, Junghyun and Mart{\'\i}nez-Ram{\'\i}rez, Marco A and Liao, Wei-Hsiang and Uhlich, Stefan and Lee, Kyogu and Mitsufuji, Yuki},
  booktitle={ICASSP 2023-2023 IEEE International Conference on Acoustics, Speech and Signal Processing (ICASSP)},
  pages={1--5},
  year={2023},
  organization={IEEE}
}

@article{li2023mert,
  title={Mert: Acoustic music understanding model with large-scale self-supervised training},
  author={Li, Yizhi and Yuan, Ruibin and Zhang, Ge and Ma, Yinghao and Chen, Xingran and Yin, Hanzhi and Xiao, Chenghao and Lin, Chenghua and Ragni, Anton and Benetos, Emmanouil and others},
  journal={arXiv preprint arXiv:2306.00107},
  year={2023}
}

@article{li2022map,
  title={Map-music2vec: A simple and effective baseline for self-supervised music audio representation learning},
  author={Li, Yizhi and Yuan, Ruibin and Zhang, Ge and Ma, Yinghao and Lin, Chenghua and Chen, Xingran and Ragni, Anton and Yin, Hanzhi and Hu, Zhijie and He, Haoyu and others},
  journal={arXiv preprint arXiv:2212.02508},
  year={2022}
}

@article{baevski2020wav2vec,
  title={wav2vec 2.0: A framework for self-supervised learning of speech representations},
  author={Baevski, Alexei and Zhou, Yuhao and Mohamed, Abdelrahman and Auli, Michael},
  journal={Advances in neural information processing systems},
  volume={33},
  pages={12449--12460},
  year={2020}
}

@article{muffin,
  title={Multi-band Frequency Reconstruction for Neural Psychoacoustic Coding},
  author={Ng, Dianwen and Zhou, Kun and Chao, Yi-Wen and Xiong, Zhiwei and Ma, Bin and Chng, Eng Siong},
  journal={arXiv preprint arXiv:2505.07235},
  year={2025}
}

@inproceedings{liu2022convnet,
  title={A convnet for the 2020s},
  author={Liu, Zhuang and Mao, Hanzi and Wu, Chao-Yuan and Feichtenhofer, Christoph and Darrell, Trevor and Xie, Saining},
  booktitle={Proceedings of the IEEE/CVF conference on computer vision and pattern recognition},
  pages={11976--11986},
  year={2022}
}

@article{dosovitskiy2020image,
  title={An image is worth 16x16 words: Transformers for image recognition at scale},
  author={Dosovitskiy, Alexey and Beyer, Lucas and Kolesnikov, Alexander and Weissenborn, Dirk and Zhai, Xiaohua and Unterthiner, Thomas and Dehghani, Mostafa and Minderer, Matthias and Heigold, Georg and Gelly, Sylvain and others},
  journal={arXiv preprint arXiv:2010.11929},
  year={2020}
}

@inproceedings{cai2023efficientvit,
  title={Efficientvit: Lightweight multi-scale attention for high-resolution dense prediction},
  author={Cai, Han and Li, Junyan and Hu, Muyan and Gan, Chuang and Han, Song},
  booktitle={Proceedings of the IEEE/CVF international conference on computer vision},
  pages={17302--17313},
  year={2023}
}

@article{afchar2025fourier,
  title={A Fourier Explanation of AI-music Artifacts},
  author={Afchar, Darius and Meseguer-Brocal, Gabriel and Akesbi, Kamil and Hennequin, Romain},
  journal={arXiv preprint arXiv:2506.19108},
  year={2025}
}

@article{kim2025segment,
  title={Segment Transformer: AI-Generated Music Detection via Music Structural Analysis},
  author={Kim, Yumin and Go, Seonghyeon},
  journal={arXiv preprint arXiv:2509.08283},
  year={2025}
}

\end{document}